\documentclass[conference]{IEEEtran}
\IEEEoverridecommandlockouts
\usepackage{cite}
\usepackage{amsmath,amssymb,amsfonts}
\usepackage{algorithmic}
\usepackage{graphicx}
\usepackage{textcomp}
\usepackage{xcolor}
\usepackage{multirow}
\usepackage{booktabs}
\usepackage{makecell}
\usepackage{tabularx}
\usepackage{soul}
\usepackage[hidelinks]{hyperref}
\usepackage[normalem]{ulem}
\soulregister\sout7

\def\BibTeX{{\rm B\kern-.05em{\sc i\kern-.025em b}\kern-.08em
    T\kern-.1667em\lower.7ex\hbox{E}\kern-.125emX}}

\begin{document}

\title{Unifying Acoustic Features and Text with Multimodal LLMs for Neurodegenerative Screening\\
}

\author{\IEEEauthorblockN{Qingfeng Zhang, Yuanxiong Guo}
\IEEEauthorblockA{\textit{Dept. of Information Systems and Cybersecurity} \\
\textit{The University of Texas at San Antonio}\\
San Antonio, TX, USA \\
\{qingfeng.zhang, yuanxiong.guo\}@utsa.edu}
\and
\IEEEauthorblockN{Yanmin Gong}
\IEEEauthorblockA{\textit{School of Engineering Medicine} \\
\textit{Texas A\&M University}\\
Houston, TX, USA \\
yanmin.gong@tamu.edu}
}

\maketitle

\begin{abstract}
Voice-based screening offers a scalable and non-invasive way to assess neurodegenerative diseases such as Alzheimer's disease (AD) and Parkinson's disease (PD), but their staging remains challenging due to the difficulty of integrating heterogeneous data. %
%
%
This paper presents NeurMLLM, an efficient multimodal generative framework for neurodegenerative disease staging. NeurMLLM first encodes the spectrograms and Mel-frequency cepstral coefficients of audio data with vision transformers and projects their representations into the embedding space of a large language model (LLM), where they are concatenated with transcript and demographic instruction tokens as a single unified sequence. %
The LLM is then instruction-tuned via Low-Rank Adaptation using task prompts to autoregressively predict a constrained label token, enabling a generative classification. %
By evaluating on the Bridge2AI-Voice dataset for fine-grained staging of AD and PD, we observe that NeurMLLM achieves strong performance, consistently outperforming classical machine learning methods and existing LLM-based approaches. The results show the high potential of multimodal LLMs in neurodegenerative disease staging, improving staging accuracy and supporting accessible deployment.  

\end{abstract}


\section{Introduction}
Neurodegenerative diseases, particularly Alzheimer’s disease (AD) and Parkinson’s disease (PD), impose a growing burden on global healthcare systems and are frequently underdiagnosed until irreversible damage has occurred \cite{wang2024expanding}. Although neuroimaging and other clinical tests can be definitive, their cost, invasiveness, and limited accessibility hinder large-scale screening. In contrast, the widespread availability of smartphones enables scalable and low-cost voice assessments outside clinical settings \cite{olson2022smartphone, wang2023applications}. Because neurodegeneration disrupts speech motor control and cognitive language planning, early impairment can manifest as measurable changes in prosody and fluency, including slowed articulation, reduced pitch variation, and irregular hesitations \cite{martinez2021ten,favaro2024unveiling}. These acoustic and linguistic cues motivate voice as a practical digital biomarker for neurodegenerative screening, where fine-grained staging is essential for timely and targeted intervention.

Large Language Models (LLMs) provide a promising foundation for building unified diagnostic interfaces by leveraging shared language representation and a consistent decision mechanism across tasks \cite{gao2025federated, yang2026fedkrso}. Recent studies have applied prompting or LLM embeddings to voice transcripts for AD or PD related inference \cite{crawford2025linguistic,castelli2025detecting}, suggesting that LLM-based decision modules can capture clinically relevant linguistic signatures. Instruction-tuned LLMs can follow task specifications expressed as prompts and produce structured outputs within a single generative framework, which is appealing for multi-condition staging scenarios.

Despite these advances, two challenges remain for practical neurodegenerative voice screening. First, many LLM-based studies are largely transcript-centric and fail to integrate acoustic representations with linguistic and demographic context in a single model \cite{crawford2025linguistic,castelli2025detecting}. Second, many LLM systems still rely on a classification head that maps pooled representations to a fixed label space, resulting in a rigid decision layer that is difficult to generalize across tasks and less aligned with token-level multimodal interactions \cite{mo2025dect, zheng2024alzheimer}. 

To address these challenges, we propose NeurMLLM, a multimodal generative framework for fine-grained neurodegenerative staging under resource-constrained settings. NeurMLLM encodes acoustic features of patient voice data, including spectrograms and Mel-frequency cepstral coefficients (MFCCs), with vision transformer (ViT)  \cite{dosovitskiy2021an}, which capture complementary time-frequency and cepstral characteristics. A projection layer aligns these acoustic embeddings with the embedding space of backbone LLM, where they are concatenated with the transcript and demographic prompt tokens. We fine-tune the model using low-rank adaptation (LoRA) \cite{hu2022lora,yeh2024navigating} and formulate the staging as constrained label-token generation, where the model is trained to generate task-specific stage tokens autoregressively.

The main contributions of this work are summarized as follows: \textbf{(1) Multimodal architecture for voice-based neurodegenerative staging.} We introduce a multimodal staging framework that integrates acoustic features, transcripts, and demographic contexts. This design enables complementary modeling of paralinguistic, linguistic, and population factors for fine-grained neurodegenerative disease staging with voice data. \textbf{(2) Instruction-tuned LLM for generative staging.} By aligning multimodal biomarkers in an instruction-tuned LLM and performing constrained label-token generation, NeurMLLM outperforms classical machine learning methods and LLM-based methods with classification head formulations, and achieves the best results on fine-grained AD and PD staging. \textbf{(3) Extensive experimental validation on fine-grained staging.} We conduct comprehensive experiments on the Bridge2AI-Voice dataset (v3.0.0) \cite{b2ai_voice2025} for multi-class AD and PD staging. Our results demonstrate that NeurMLLM consistently outperforms traditional machine learning methods and LLM-based methods with classification heads in detecting different AD and PD stages.

\section{Related work}

\subsection{Voice Representation Learning}

The advance of deep learning has substantially improved pathological speech analysis by enabling representation learning from raw audio \cite{shi2023speech,ding2024speech}. Early studies have primarily used convolutional neural networks (CNNs) to learn local spectral and temporal structure from spectrogram representations for neurodegenerative screening  \cite{warnita2018detecting}, while long short-term memory (LSTM) networks have been explored to model longer temporal dependencies in acoustic features \cite{xue2021detection}. More recently, Ilias et al. \cite{ilias2023detecting} have demonstrated that ViT encoding of spectrograms and MFCCs representations can effectively capture subtle acoustic deviations associated with dementia. Multimodal fusion combining acoustic features with automatic speech recognition (ASR) transcripts has shown promise through techniques such as late fusion and cross-modal attention \cite{shao25_interspeech, ilias2023detecting, abid2025speechhgt, escobar2025synchronous}. However, these pipelines lack a unified framework that jointly models acoustic features, transcripts, and demographic context within a single architecture.

\subsection{LLMs for Neurodegenerative Screening}





LLMs have emerged as a flexible tool for neurodegenerative screening, but most existing work remains transcript-centric \cite{crawford2025linguistic, castelli2025detecting}. On the modeling side, existing studies often use a classification head that maps pooled LLM representations to a fixed label space \cite{mo2025dect, zheng2024alzheimer}. Recent evidence suggests that generative classifiers can outperform discriminative counterparts in low-data regimes \cite{kasa2025generative}. Consistent with this finding, Taherinezhad et al. \cite{taherinezhad2026large} have shown that token-level fine-tuning generally outperforms classification head-based approaches for cognitive impairment detection. In parallel, Casu et al. \cite{casu2025integrating} combine fine-tuned LLM embeddings with acoustic features, but still rely on an external classifier. These limitations motivate a multimodal framework that jointly aligns acoustic, linguistic, and demographic information within an LLM and reformulates neurodegenerative disease staging as a constrained label-token generation problem.

\section{Method}
\begin{figure*}[t]
\centering
\includegraphics[width=\linewidth]{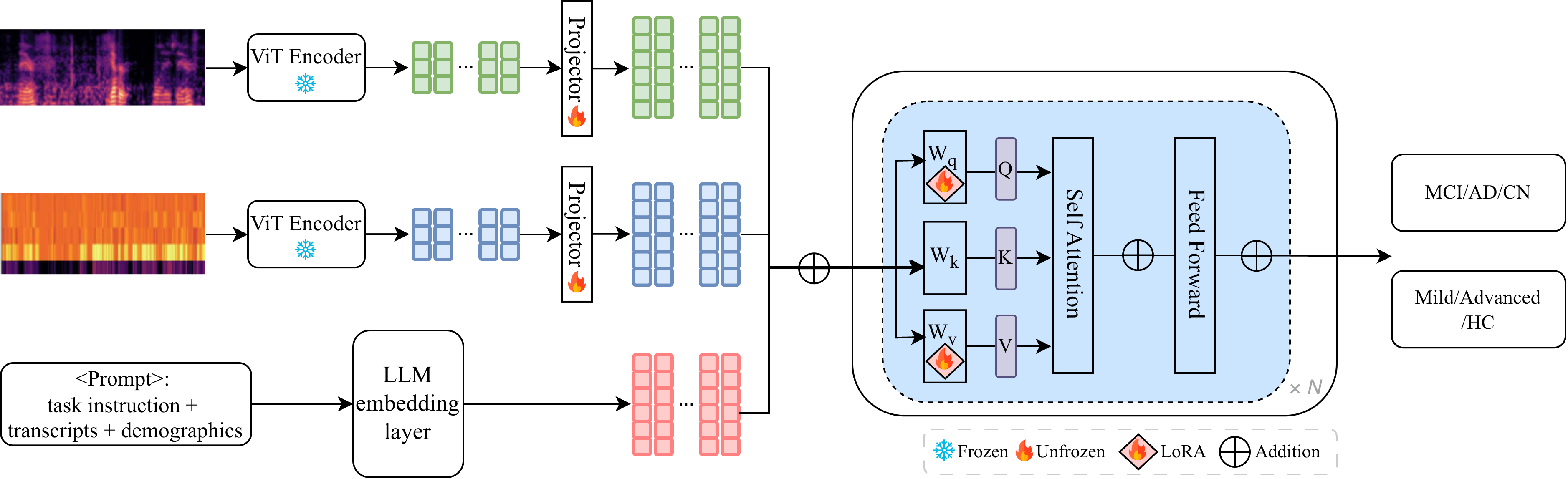}
\caption{Model architecture of NeurMLLM. Text embeddings from prompts, along with acoustic embeddings from spectrograms and MFCCs, are sequentialized as input for the backbone LLM.}
\label{fig:NeurMLLM}
\end{figure*}

The NeurMLLM framework is illustrated in Fig. \ref{fig:NeurMLLM}. We propose a multimodal generative architecture for fine-grained neurodegenerative staging, fusing acoustic features including spectrograms and MFCCs with linguistic context. We systematically compare this generative label-token paradigm with traditional machine learning and classification head approaches on the Bridge2AI-Voice dataset.

\subsection{Datasets}\label{AA}

The Bridge2AI-Voice dataset (v3.0.0) is designed with a strong emphasis on privacy, providing derived acoustic features instead of raw audio recordings. It offers a multimodal representation that includes spectrograms and MFCCs as complementary acoustic features, where spectrograms encode time–frequency energy distributions and MFCCs capture compact representations of the spectral envelope associated with vocal tract characteristics. In addition, the dataset includes aligned speech transcripts and basic demographic attributes (e.g., age and gender). This combination of modalities enables integrated analysis of paralinguistic vocal patterns, linguistic content, and demographic context.

We include multiple acoustic tasks available in Bridge2AI-Voice to capture complementary cognitive--linguistic and motor--prosodic cues. Specifically, we use \textit{Cinderella-story} (narrative recall), \textit{Picture description} (spontaneous description), \textit{Word-color Stroop} (inhibitory control under cognitive load), \textit{Productive vocabulary} (lexical retrieval), and \textit{Random item generation} (verbal fluency). For each participant in the dataset, we aggregate all available samples from these acoustic tasks under a participant-level protocol and concatenate the corresponding task-tagged transcripts while pairing them with the aligned spectrogram and MFCC features.

We aim to perform fine-grained, participant-level disease staging from multimodal voice evidence. Given a participant's acoustic features, aligned transcript, and demographic attributes (age and gender), the model predicts a stage label for AD or PD. We constructed two task-specific staging cohorts for AD and PD, totaling 156 participants for AD and 167 for PD. For the AD task, participants were stratified into Cognitive Normal (CN, $n=83$), Mild Cognitive Impairment (MCI, $n=44$), and Alzheimer’s Disease (AD, $n=29$) to target stage detection. For the PD task, participants were grouped by Hoehn and Yahr stages into Stage 1–2 (Early, $n=34$), Stage 3–5 (Advanced, $n=50$), and Healthy Control (HC, $n=83$). Healthy controls are included in both tasks as the reference class. Table \ref{tab:dataset_stats_neurmllm} summarizes the statistics of the constructed cohorts. 

\begin{table}[htbp]
\centering
\caption{Dataset Statistics}
\label{tab:dataset_stats_neurmllm}
\begin{tabular}{c c c c}
\toprule
\textbf{Task} & \textbf{Class} & \textbf{\#Participants} & \textbf{\#Acoustic Samples} \\
\midrule
\multirowcell{3}{\textbf{AD}} 
& MCI & 44 & 586 \\
& AD  & 29 & 356 \\
& CN  & 83 & 218 \\
\midrule
\multirowcell{3}{\textbf{PD}} 
& Mild & 34 & 422 \\
& Advanced & 50 & 505 \\
& HC & 83 & 218 \\
\bottomrule
\end{tabular}
\end{table}


\subsection{Model Architecture}

\noindent\textbf{Inputs.}
Each participant consists of two acoustic modalities and one text instruction. The acoustic input includes a spectrogram map $S$ and an MFCC map $M$. The textual input is an instruction string $T$ built from a task-specific template that includes the transcript and demographic information.

\noindent\textbf{Acoustic Encoders with Projectors.}
We employ two modality-specific ViT encoders to extract embedding representations from the acoustic maps:
\begin{equation}
    Z_s = E_{\mathrm{spec}}(S), \qquad Z_m = E_{\mathrm{mfcc}}(M),
\end{equation}
where $E_{\mathrm{spec}}(\cdot)$ and $E_{\mathrm{mfcc}}(\cdot)$ denote the ViT feature extractors for spectrogram and MFCC inputs, respectively, and $Z_s \in \mathbb{R}^{L_s \times d_v}$ and $Z_m \in \mathbb{R}^{L_m \times d_v}$ are sequences of acoustic embeddings. Since the ViT embedding dimension $d_v$ generally differs from the LLM hidden size $d_\ell$, we use two lightweight linear layers as the projectors:
\begin{equation}
    \hat{Z}_s = P_{\mathrm{spec}}(Z_s), \qquad \hat{Z}_m = P_{\mathrm{mfcc}}(Z_m),
\end{equation}
where $P_{\mathrm{spec}}(\cdot)$ and $P_{\mathrm{mfcc}}(\cdot)$ are learnable linear projectors and $\hat{Z}_s \in \mathbb{R}^{L_s \times d_\ell}$ and $\hat{Z}_m \in \mathbb{R}^{L_m \times d_\ell}$.

\noindent\textbf{Text Embeddings.}
The instruction text $T$ is tokenized and embedded by the LLM embedding layer:
\begin{equation}
    X = \mathrm{Embed}(T), \qquad X \in \mathbb{R}^{L_t \times d_\ell}.
\end{equation}
where $\mathrm{Embed}(\cdot)$ denotes the LLM input embedding layer.

\noindent\textbf{Concatenation and Self-Attention Fusion.}
Following recent multimodal LLM practice, we build a single unified token sequence by concatenating special modality tags and projected acoustic embeddings with the instruction embeddings, enabling cross-modal interaction through the LLM self-attention:
\begin{equation}
U = \left[\mathrm{[CLS]}, \mathrm{[SPEC]}, \hat{Z}_s, \mathrm{[MFCC]}, \hat{Z}_m, \mathrm{[TEXT]},X \right],
\end{equation}
where $\mathrm{[CLS]}$ is a global context token;  $\mathrm{[SPEC]}$, $\mathrm{[MFCC]}$, and $\mathrm{[TEXT]}$ are learnable modality tags; $\hat{Z}_s \in \mathbb{R}^{L_s \times d_\ell}$ and $\hat{Z}_m \in \mathbb{R}^{L_m \times d_\ell}$ denote the projected spectrograms and MFCCs sequences, and $X \in \mathbb{R}^{L_t \times d_\ell}$ is the text embedding sequence. Given the token sequence $U \in \mathbb{R}^{L \times d_\ell}$, each transformer layer performs multi-head self-attention:
\begin{equation}
    \mathrm{Attn}(U) = \mathrm{Softmax}\!\left(\frac{QK^\top}{\sqrt{d_h}}\right)V,
\end{equation}
with linear projections
\begin{equation}
    Q = UW_q,\quad K = U W_k,\quad V = UW_v,
\end{equation}
where $W_q, W_k, W_v \in \mathbb{R}^{d_\ell \times d_\ell}$ are learnable projection matrices and $d_h$ denotes the head dimension in multi-head attention. This mechanism enables tokens from different modalities to attend to each other within the same sequence.


\subsection{Fine-tuning}
To enable efficient adaptation under resource constraints, we fine-tune the LLM using LoRA, updating only a small set of rank-decomposed matrices while keeping the majority of backbone weights frozen. 

\noindent\textbf{LoRA adaptation.} For a base model weight matrix $W \in \mathbb{R}^{d \times d}$, LoRA updates it as follows: 
\begin{equation}
W' = W + \Delta W, \qquad \Delta W = \frac{\alpha}{r}B A,
\end{equation}
where $B\in\mathbb{R}^{d\times r}$ and $A\in\mathbb{R}^{r\times d}$ are trainable low-rank matrices and $r \ll d$ is the rank. We apply LoRA to the attention \emph{query} and \emph{value} projection matrices ($W_q$ and $W_v$) of the transformer block, while keeping other weights unchanged. This design provides a good trade-off between preserving the backbone's pre-trained knowledge and adapting to the downstream staging task with minimal additional trainable parameters, which is particularly important under limited data and compute budgets.

\subsection{Optimization Objective}

We formulate the staging as a constrained label-token generation task. Given the multimodal input sequence $U$, the LLM produces next-token logits. We compute the probability over a predefined label set $\mathcal{Y}$ using the logits corresponding to the label token:
\begin{equation}
    p_\theta(y \mid U) = \frac{\exp(o_y)}{\sum_{y' \in \mathcal{Y}} \exp(o_{y'})},
\end{equation}
where $o_y$ denotes the logit for label token $y$. The model is trained with cross-entropy loss on the ground-truth label $y^*$,
\begin{equation}
    \mathcal{L} = -\log p_\theta(y^* \mid U),
\end{equation}
and the predicted label is given by:
\begin{equation}
    \hat{y} = \arg\max_{y \in \mathcal{Y}} p_\theta(y \mid U).
\end{equation}

\begin{figure}[t]
\centering
\includegraphics[width=\linewidth]{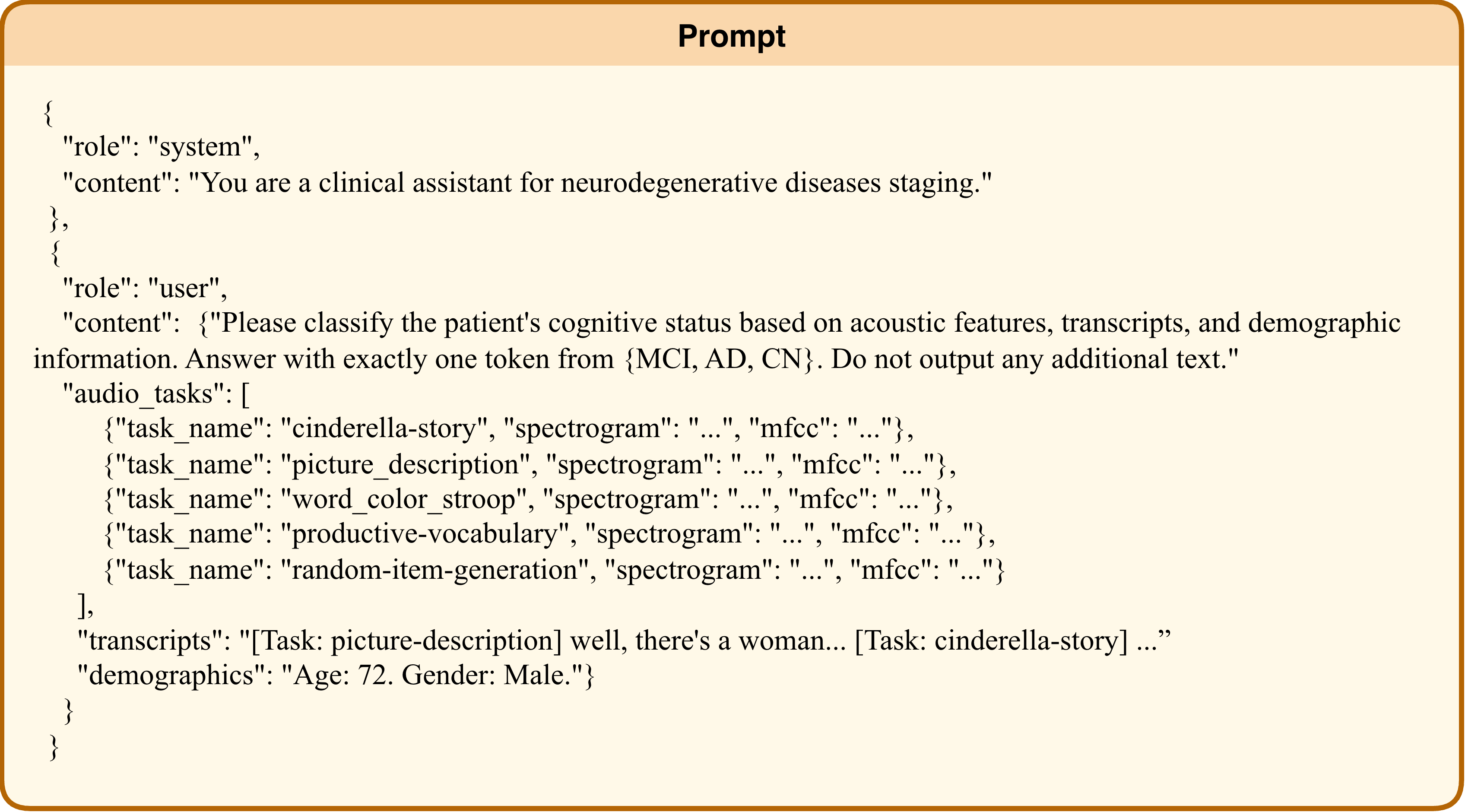}
\caption{Example prompt template and structured multimodal input. }
\label{fig:prompt}
\end{figure}

\subsection{Prompts}
As shown in Fig. \ref{fig:prompt}, we use a chat-style prompt to unify multimodal evidence into a single instruction-following input. Each instance comprises two messages: a \textit{system} message that defines the model's role and a \textit{user} message that includes (i) a task instruction constraining the output space and (ii) a structured input that includes acoustic features, transcripts, and demographics. The instruction requests exactly one categorical label, enforcing constrained label-token generation rather than a task-specific classification head. The AD and PD tasks share the same schema but use different task identifiers and label sets: AD staging uses \{\texttt{MCI}, \texttt{AD}, \texttt{CN}\}, whereas PD staging uses \{\texttt{Mild}, \texttt{Advanced}, \texttt{HC}\}. This specification provides a unified generative interface while supporting task-specific staging criteria.

\subsection{Experiment Settings}

\subsubsection{Training and Evaluation} 

We adopt Llama-3.2-3B-Instruct as the backbone and apply LoRA to the \emph{query} and \emph{value} projection matrices ($W_q$ and $W_v$) with $r=8$, $\alpha=16$, and dropout $p=0.05$. We use batch size 16 to fine-tune the LoRA matrices for 3 epochs on an RTX A6000 with \texttt{bfloat16}. We report participant-level performance by aggregating per-sample predictions for each participant under a stratified 60/20/20 train/val/test split. Metrics include macro-AUROC, Accuracy (Acc.), macro-F1, and macro-Recall to account for class imbalance in the multi-class staging. 

\subsubsection{Baseline}
For comparison, we implement several baselines under the same participant-level protocol. 
We include a classical machine learning baseline, namely \textbf{LR}, which concatenates acoustic representations, transcripts, and demographic attributes at the feature level and trains a multinomial logistic regression classifier. 
We also evaluate a cross-attention fusion baseline, referred to as \textbf{CrossAttn}, which fuses acoustic and textual tokens via cross-attention and predicts labels with a linear classifier. 
In addition, we consider an LLM-based method with a classification head, denoted as \textbf{ClsHead}, which predicts class logits using a task-specific classification head trained with cross-entropy. 
Finally, we include the method of Casu et al. \cite{casu2025integrating}, which fine-tunes an LLM on transcript-based prompts and leverages the resulting hidden-state representations as linguistic features, which are then fused with acoustic features for classification, referred to as \textbf{LLM-A-X}.

\begin{table*}[t]
\centering
\caption{Performance Comparison on AD and PD Tasks. Results are reported as mean $\pm$ standard deviation over five runs. The best performance for each metric is highlighted in \textbf{bold}.}
\label{tab:performance}
\begin{tabular}{lcccccccc}
\toprule
& \multicolumn{4}{c}{\textbf{AD}} & \multicolumn{4}{c}{\textbf{PD}} \\
\cmidrule(lr){2-5} \cmidrule(lr){6-9}
\textbf{Baseline} &
macro-AUROC & Acc. & macro-F1 & macro-Recall &
macro-AUROC & Acc. & macro-F1 & macro-Recall \\
\midrule
LR
& 0.587$\pm$0.035 & 0.496$\pm$0.054 & 0.525$\pm$0.054 & 0.517$\pm$0.046
& 0.657$\pm$0.042 & 0.476$\pm$0.046 & 0.461$\pm$0.037 & 0.543$\pm$0.056\\
CrossAttn
& 0.638$\pm$0.032 & 0.526$\pm$0.050 & 0.524$\pm$0.058 & 0.480$\pm$0.064
& 0.692$\pm$0.035 & 0.497$\pm$0.044 & 0.471$\pm$0.033 & 0.566$\pm$0.047\\
ClsHead
& 0.852$\pm$0.036 & 0.740$\pm$0.031 & 0.662$\pm$0.028 & 0.709$\pm$0.031
& 0.823$\pm$0.037 & 0.658$\pm$0.032 & 0.504$\pm$0.045 & 0.607$\pm$0.037 \\
LLM-A-X
& 0.829$\pm$0.033 & 0.716$\pm$0.035 & 0.608$\pm$0.056 & 0.616$\pm$0.033
& 0.717$\pm$0.056 & 0.532$\pm$0.048 & 0.489$\pm$0.055 & 0.559$\pm$0.040\\
\midrule
\textbf{NeurMLLM}
& \textbf{0.917$\pm$0.054} & \textbf{0.823$\pm$0.031} & \textbf{0.757$\pm$0.028} & \textbf{0.748$\pm$0.036}
& \textbf{0.872$\pm$0.035} & \textbf{0.735$\pm$0.057} & \textbf{0.537$\pm$0.041} & \textbf{0.648$\pm$0.047} \\
\bottomrule
\end{tabular}
\end{table*}

\begin{table*}[t]
\centering
\caption{Modality Contribution Analysis. Results are reported as mean $\pm$ standard deviation over five runs. The best performance for each metric is highlighted in \textbf{bold}.}
\label{tab:modality_ablation}
\begin{tabularx}{\textwidth}{l *{8}{>{\centering\arraybackslash}X}}
\toprule
& \multicolumn{4}{c}{\textbf{AD}} & \multicolumn{4}{c}{\textbf{PD}} \\
\cmidrule(lr){2-5} \cmidrule(lr){6-9}
\textbf{Setting} & macro-AUROC & Acc. & macro-F1 & macro-Recall & macro-AUROC & Acc. & macro-F1 & macro-Recall \\
\midrule
Audio only
& 0.903$\pm$0.026 & 0.794$\pm$0.023 & 0.731$\pm$0.029 & 0.714$\pm$0.034
& 0.864$\pm$0.027 & 0.676$\pm$0.039 & 0.481$\pm$0.021 & 0.564$\pm$0.031 \\
Text only
& 0.647$\pm$0.055 & 0.556$\pm$0.047 & 0.230$\pm$0.056 & 0.333$\pm$0.037
& 0.826$\pm$0.053 & 0.529$\pm$0.062 & 0.222$\pm$0.044 & 0.333$\pm$0.046 \\
\midrule
\textbf{NeurMLLM}
& \textbf{0.917$\pm$0.054} & \textbf{0.823$\pm$0.031} & \textbf{0.757$\pm$0.028} & \textbf{0.748$\pm$0.036}
& \textbf{0.872$\pm$0.035} & \textbf{0.735$\pm$0.057} & \textbf{0.537$\pm$0.041} & \textbf{0.648$\pm$0.047} \\
\bottomrule
\end{tabularx}
\end{table*}

\begin{table*}[t]
\centering
\caption{Impact of LLM Backbones and Instruction Alignment. Results are reported as mean $\pm$ standard deviation over five runs.  The best performance for each metric is highlighted in \textbf{bold}.}
\label{tab:llm_ablation}
\begin{tabularx}{\textwidth}{l *{8}{>{\centering\arraybackslash}X}}
\toprule
& \multicolumn{4}{c}{\textbf{AD}} & \multicolumn{4}{c}{\textbf{PD}} \\
\cmidrule(lr){2-5} \cmidrule(lr){6-9}
\textbf{Backbone} & macro-AUROC & Acc. & macro-F1 & macro-Recall & macro-AUROC & Acc. & macro-F1 & macro-Recall\\
\midrule
Qwen2.5-3B-Instruct & 0.858$\pm$0.042 & 0.676$\pm$0.055 & 0.559$\pm$0.041 & 0.559$\pm$0.031 & 0.853$\pm$0.041 & 0.647$\pm$0.048 & 0.472$\pm$0.059 & 0.569$\pm$0.051\\
Llama-3.2-3B        & 0.903$\pm$0.036 & 0.794$\pm$0.029 & 0.696$\pm$0.052 & 0.692$\pm$0.058 & 0.869$\pm$0.060 & 0.726$\pm$0.039 & \textbf{0.620$\pm$0.055} & \textbf{0.689$\pm$0.048}\\
\midrule
\textbf{NeurMLLM}
& \textbf{0.917$\pm$0.054} & \textbf{0.823$\pm$0.031} & \textbf{0.757$\pm$0.028} & \textbf{0.748$\pm$0.036}
& \textbf{0.872$\pm$0.035} & \textbf{0.735$\pm$0.057} & 0.537$\pm$0.041 & 0.648$\pm$0.047 \\
\bottomrule
\end{tabularx}
\end{table*}

\section{Results}

\subsection{Overall Performance Comparison}
Table~\ref{tab:performance} summarizes the performance of different methods on AD and PD staging tasks. NeurMLLM achieves the strongest overall results across both diseases, indicating that constrained label-token generation provides a more effective formulation for stage prediction than conventional classification formulation. In particular, compared with \textit{ClsHead}, NeurMLLM improves macro-AUROC from 0.852 to 0.917 and Accuracy from 0.740 to 0.823, while also achieving the best macro-F1 (0.757) and macro-Recall (0.748). These gains suggest that directly optimizing the probability of the correct label token can better align multimodal evidence with discrete clinical stages than mapping the representation to a fixed logit vector through a linear head. On PD staging, NeurMLLM yields the best macro-AUROC (0.872) and Accuracy (0.735), and attains the highest macro-Recall (0.648), highlighting improved sensitivity when separating mild (Stage~1--2) from advanced (Stage~3--5) cases. Compared with traditional LR and cross-attention fusion methods, our approach shows clear advantages on AD and maintains competitive performance on PD, supporting the robustness of the proposed multimodal generative formulation for fine-grained staging under limited data.

\subsection{Modality Contribution Analysis}
To quantify the contributions of acoustic and linguistic inputs, we evaluate single-modality and multimodal settings as shown in Table~\ref{tab:modality_ablation}. The \textit{Audio only} setting achieves strong performance, indicating that paralinguistic cues captured by spectrograms and MFCCs are effective for separating disease stages. In contrast, \textit{Text only} performs poorly on classification metrics, even though AD macro-AUROC appears relatively high, suggesting that transcripts alone are insufficient for fine-grained staging. Importantly, multimodal fusion consistently yields the best results across all metrics: our model improves AD staging Accuracy to 0.823 and PD staging Accuracy to 0.735, while increasing PD macro-Recall from 0.564 to 0.648. These gains imply that linguistic information adds complementary context beyond acoustic biomarkers, leading to more reliable staging.

\subsection{Impact of LLM Backbones and Instruction Alignment}
We further investigate how different LLM backbones and instruction alignment affect multimodal staging performance. As shown in Table~\ref{tab:llm_ablation}, we evaluate \textit{Qwen2.5-3B-Instruct}, \textit{Llama-3.2-3B}, and \textit{Llama-3.2-3B-Instruct} as the backbone used in our framework. On AD staging, instruction tuning is clearly beneficial: compared with the base Llama model, \textit{Llama-3.2-3B-Instruct} improves Accuracy by 3.7\% (0.794 vs.\ 0.823) and macro-F1 by 8.8\% (0.696 vs.\ 0.757), indicating better alignment to the staging objective. Across similarly sized backbones, the Llama family also outperforms Qwen on AD (0.823 vs. 0.676 Accuracy). On PD staging, Llama models achieve the highest Accuracy (0.735), outperforming Qwen (0.647); however, the base Llama attains higher macro-F1 than the instruct model (0.620 vs. 0.537). This suggests that macro-averaged metrics are important for evaluating fine-grained staging under class imbalance.

\section{Discussion}
This study demonstrates the effectiveness of a multimodal framework in enhancing voice-based neurodegenerative staging. A key finding is that constrained label-token generation outperforms classification-head approaches in our experiments. One possible explanation is that classification heads are randomly initialized and must be learned from scratch, which is challenging given our small cohorts. In contrast, generative decoding reuses the pre-trained output projection without introducing additional parameters, preserving consistency with the pretraining objective of the LLM and instruction-following capability. 
%
%
Ablation studies confirm that while derived acoustic features remain the dominant signal source, integrating transcripts and demographics provides critical complementary information, improving recall in borderline cases. This indicates that linguistic cues and context help resolve ambiguities that acoustics alone may miss, though reliance on text alone remains insufficient due to its limited classification power in isolation.

Despite promising results, several limitations warrant consideration. First, the small cohort size introduces performance variance; findings should thus be interpreted as encouraging signals requiring validation on larger, multi-institutional datasets. Expanding the dataset would also help assess the robustness and generalizability of the proposed framework across different conditions. Second, while we adopt constrained label-token generation for classification, the generative interface could be further exploited beyond discrete labels, and the underlying mechanisms that contribute to its effectiveness remain to be further investigated.

Future work may extend the output space to additional clinically meaningful generation tasks, such as brief participant-level summaries and explanations grounded in acoustic and linguistic cues, while also incorporating additional modalities, such as clinical records or neuroimaging data, to further improve staging performance. It would also be valuable to explore how these extensions generalize across datasets and patient populations.

\section{Conclusion}
We presented NeurMLLM, a multimodal instruction-tuned framework for voice-based neurodegenerative disease screening. By encoding derived acoustic features with ViT and integrating transcript and demographic context within an instruction prompt, NeurMLLM enables a generative decision mechanism via constrained label-token prediction. Experiments on participant-level evaluation for AD and PD demonstrate that this formulation consistently outperforms LLM-based methods with classification heads and classical machine learning baselines, with notable improvements in macro-averaged metrics that reflect improved handling of minority classes. 

Overall, these results suggest that reframing staging as label token generation, together with parameter-efficient instruction tuning, provides a simple and effective alternative to classification heads. More importantly, by integrating acoustic, linguistic, and demographic information within a unified framework, NeurMLLM offers a flexible and scalable solution for multimodal voice-based neurodegenerative screening.

\section{Acknowledgment}
Q. Zhang and Y. Guo were partially supported by a seed grant from UT San Antonio Office of Research and Innovation and NSF Grant CNS-2106761. Y. Gong was partially supported by NSF Grant CNS-2611068.

\bibliographystyle{IEEEtran}
\bibliography{references}

\end{document}